\documentclass[twoside,twocolumn,9pt]{article}
\usepackage{extsizes}
\usepackage[super,sort&compress,comma]{natbib} 
\usepackage[version=3]{mhchem}
\usepackage[left=1.5cm, right=1.5cm, top=1.785cm, bottom=2.0cm]{geometry}
\usepackage{balance}
\usepackage{mathptmx}
\usepackage{sectsty}
\usepackage{graphicx} 
\usepackage{lastpage}
\usepackage[format=plain,justification=justified,singlelinecheck=false,font={stretch=1.125,small,sf},labelfont=bf,labelsep=space]{caption}
\usepackage{float}
\usepackage{fancyhdr}
\usepackage{fnpos}
\usepackage[english]{babel}
\addto{\captionsenglish}{
  
}
\usepackage{array}
\usepackage{droidsans}
\usepackage{charter}
\usepackage[T1]{fontenc}
\usepackage[usenames,dvipsnames]{xcolor}
\usepackage{setspace}
\usepackage[compact]{titlesec}
\usepackage{hyperref}

\usepackage{epstopdf}

\usepackage{orcidlink}

\definecolor{cream}{RGB}{222,217,201}

\begin{document}

\pagestyle{fancy}
\thispagestyle{plain}
\fancypagestyle{plain}{

\renewcommand{\headrulewidth}{0pt}
}

\makeFNbottom
\makeatletter
\renewcommand\LARGE{\@setfontsize\LARGE{15pt}{17}}
\renewcommand\Large{\@setfontsize\Large{12pt}{14}}
\renewcommand\large{\@setfontsize\large{10pt}{12}}
\renewcommand\footnotesize{\@setfontsize\footnotesize{7pt}{10}}
\makeatother

\renewcommand{\thefootnote}{\fnsymbol{footnote}}
\renewcommand\footnoterule{\vspace*{1pt}
\color{cream}\hrule width 3.5in height 0.4pt \color{black}\vspace*{5pt}} 
\setcounter{secnumdepth}{5}

\makeatletter 
\renewcommand\@biblabel[1]{#1}            
\renewcommand\@makefntext[1]
{\noindent\makebox[0pt][r]{\@thefnmark\,}#1}
\makeatother 
\renewcommand{\figurename}{\small{Fig.}~}
\sectionfont{\sffamily\Large}
\subsectionfont{\normalsize}
\subsubsectionfont{\bf}
\setstretch{1.125} 
\setlength{\skip\footins}{0.8cm}
\setlength{\footnotesep}{0.25cm}
\setlength{\jot}{10pt}
\titlespacing*{\section}{0pt}{4pt}{4pt}
\titlespacing*{\subsection}{0pt}{15pt}{1pt}

\makeatletter 
\newlength{\figrulesep} 
\setlength{\figrulesep}{0.5\textfloatsep} 

\newcommand{\topfigrule}{\vspace*{-1pt}
\noindent{\color{cream}\rule[-\figrulesep]{\columnwidth}{1.5pt}} }

\newcommand{\botfigrule}{\vspace*{-2pt}
\noindent{\color{cream}\rule[\figrulesep]{\columnwidth}{1.5pt}} }

\newcommand{\dblfigrule}{\vspace*{-1pt}
\noindent{\color{cream}\rule[-\figrulesep]{\textwidth}{1.5pt}} }

\makeatother

\twocolumn[
  \begin{@twocolumnfalse}
{\includegraphics[height=30pt]{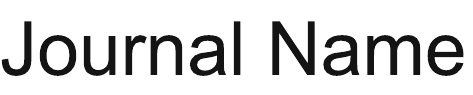}\hfill\raisebox{0pt}[0pt][0pt]{\includegraphics[height=55pt]{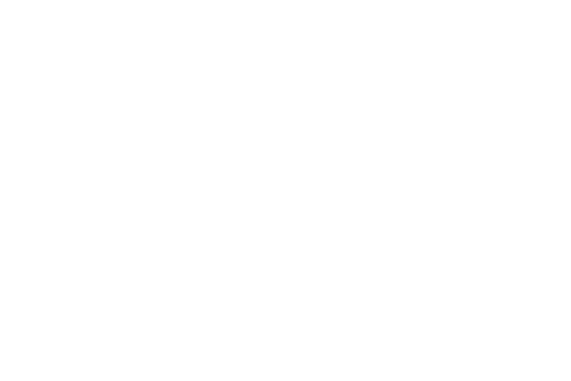}}\\[1ex]
\includegraphics[width=18.5cm]{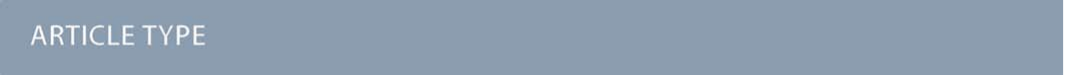}}\par
\vspace{1em}
\sffamily
\begin{tabular}{m{4.5cm} p{13.5cm} }

\includegraphics{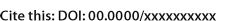} & \noindent\LARGE{\textbf{Wavelength-Resolved Photoinduced Spin Polarization in a Broad Optical Range for a Porphyrin–Quinone System}} \\
\vspace{0.3cm} & \vspace{0.3cm} \\

  & \noindent\large{Liubov Chuchkova \orcidlink{0000-0003-0763-1385}\textit{$^{a,b,c}$}, Oleg Tretiak\orcidlink{0000-0002-7667-2933},$^{\ast}$\textit{$^{a,b}$}, Kirill F. Sheberstov\orcidlink{0000-0002-3520-6258}\textit{$^{d}$}, Danila A. Barskiy\textit{$^{e}$}, Raphael Kircher\textit{$^{a,b}$}, and Dmitry Budker\orcidlink{0000-0002-7356-4814}\textit{$^{a,b,c,f}$}} \\

\includegraphics{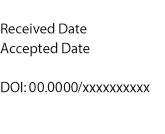} & \noindent\normalsize{Photochemically induced dynamic nuclear polarization (photo-CIDNP) in liquid-state donor–acceptor systems is typically studied at a limited number of excitation wavelengths, leaving its spectral dependence incompletely characterized. Understanding the wavelength dependence of photo-CIDNP is important both for elucidating the underlying spin-chemical mechanisms and for optimizing hyperpolarization strategies in chemically and biologically relevant molecular systems. Here, we investigate wavelength-resolved photo-CIDNP in a tetraphenylporphyrin–1,4-benzoquinone donor–acceptor system over the 350–800\,nm  spectral range. Photon-flux-normalized CIDNP amplitudes were measured using both a tunable laser system and a broadband xenon lamp equipped with interchangeable 10\,nm  interference filters.
The CIDNP response exhibits a non-monotonic dependence on excitation wavelength. Pronounced hyperpolarization is observed near 350\,nm  and in the 500–550\,nm  region, whereas excitation within the strongly absorbing 400–450\,nm  range results in a substantially reduced CIDNP response. Comparison with the UV–Vis absorption spectrum demonstrates that photo-CIDNP efficiency is not governed solely by optical absorption and reflects wavelength-dependent photophysical processes.
After normalization to the excitation photon flux, lamp- and laser-based measurements yield consistent CIDNP results, validating broadband filtered excitation as a reliable and experimentally accessible approach for wavelength-resolved photo-CIDNP studies. These results establish excitation wavelength as an independent control parameter for liquid-state photo-CIDNP and provide a framework for systematic investigations of wavelength-dependent spin hyperpolarization.} \\

\end{tabular}

 \end{@twocolumnfalse} \vspace{0.6cm}

  ]

\renewcommand*\rmdefault{bch}\normalfont\upshape
\rmfamily
\section*{}
\vspace{-1cm}

\footnotetext{\textit{$^{a}$~Institute of Physics, Johannes Gutenberg University of Mainz, 55099 Mainz, Germany.}}
\footnotetext{\textit{$^{b}$~Helmholtz Institute Mainz, 55099 Mainz, Germany.}}
\footnotetext{\textit{$^{c}$~GSI Helmholtzzentrum für Schwerionenforschung, 64291 Darmstadt, Germany.}}
\footnotetext{\textit{$^{d}$~Chimie Physique et Chimie du Vivant (CPCV, UMR 8228), Département de Chimie, École Normale Supérieure, PSL University, Sorbonne Université, 75005 Paris, France.}}
\footnotetext{\textit{$^{e}$~Department of Chemistry and Physics, University of Miami, Frost Institute for Chemistry and Molecular Science, 1201 Memorial Dr., Coral Gables, FL 33146, USA.}}
\footnotetext{\textit{$^{f}$~Department of Physics, University of California, Berkeley, CA 94720, USA.}}
\footnotetext{\textit{$^{\ast}$~E-mail: \href{mailto:oleg.tretiak@uni-mainz.de}
{oleg.tretiak@uni-mainz.de}}}

\section{Introduction}

Photochemically induced dynamic nuclear polarization (photo-CIDNP) in liquids is a widely used probe of photochemical electron-transfer reactions. Since its discovery, photo-CIDNP has provided information on the formation and evolution of radicals that is difficult to obtain otherwise \cite{KapteinOosterhoff1969_CIDNP_II,SteinerUlrich1989_MagneticFieldEffects,ClossForbesNorris1987_SpinPolarizedEPR}.

To investigate these processes 

under controlled conditions, well-defined donor–acceptor model systems are commonly used. Porphyrin–quinone pairs, in particular, have been extensively studied as benchmark systems for photoinduced electron transfer, charge separation, and spin-selective recombination because their photophysical and redox properties are well characterized \cite{Wasielewski1992_ArtificialPhotosynthesis,
Gust1991_ProtonAssistedPET,
Maruyama1986_CIDNPPorphyrinPhenolQuinone,
sheberstov2021photochemically,
chuchkova2023magnetometer}.
Beyond its role as a mechanistic probe of radical-pair reactions,
photo-CIDNP has attracted considerable interest as a
hyperpolarization technique capable of enhancing NMR sensitivity
under ambient conditions. In favorable systems, appropriately chosen
photosensitizers enable hypersensitive NMR data collection at
low-micromolar analyte concentrations
\cite{OkunoCavagnero2016_Fluorescein}.

Photo-CIDNP has also been successfully applied to aromatic amino
acids, peptides, proteins, and photosynthetic reaction centers
\cite{HoreBroadhurst1993_PhotoCIDNP,
Kuhn2013_PhotoCIDNPAminoAcidsProteins,
Kaptein1978_PhotoCIDNPProteinSurface,
Daviso2009_ElectronicStructurePrimaryDonor}.

More recently, photo-CIDNP-compatible libraries containing hundreds
of small molecules have enabled ultrafast fragment screening,
micromolar detection on benchtop NMR spectrometers, and rapid
characterization of protein--ligand interactions
\cite{Torres2023_UltrafastFragmentScreening,
Stadler2023_BenchtopFragmentScreening,
Butikofer2024_ProteinLigandAffinityPhotoCIDNP}.

Photo-CIDNP has also been demonstrated for nucleic-acid bases and
nucleotides
\cite{Kaptein1979_PhotoCIDNPNucleicAcids}.

The magnetic-field dependence of CIDNP provides an additional
mechanistic probe, because the external field modifies spin evolution
in multinuclear radical pairs and hence the resulting nuclear
polarization
\cite{Ivanov2003_FieldDependenceCIDNP}.

A better understanding of the factors governing CIDNP efficiency is
therefore important not only for fundamental spin chemistry but also
for the development and optimization of hyperpolarization approaches
for chemically and biologically relevant systems
\cite{MorozovaIvanov2019_TimeResolvedCIDNPBiomolecules}.

In most liquid-state photo-CIDNP experiments, excitation is performed at a single wavelength, typically selected near an absorption maximum of the photosensitizer and often determined by the available light source. As a result, the influence of excitation wavelength on CIDNP efficiency is rarely investigated systematically. 
However, excitation within the Soret and Q bands initially accesses
different electronic manifolds of the porphyrin, which are followed
by competing internal conversion, fluorescence, intersystem crossing,
and electron-transfer pathways
\cite{Gouterman1961_SpectraPorphyrins,
Baskin2002_UltrafastDynamicsTPP,
Gust1991_ProtonAssistedPET}.
Excitation wavelength may therefore influence radical-pair formation
and the resulting nuclear spin polarization and should be regarded as
a potentially important experimental parameter governing
photo-CIDNP efficiency.

Evidence that nuclear spin polarization can depend sensitively on optical excitation conditions has been reported in solid-state photo-CIDNP studies under magic-angle spinning \cite{Daviso2009_ElectronicStructurePrimaryDonor,
Bode2013_SolidStatePhotoCIDNP,
Matysik2022_PhotoCIDNPSolidState}. These systems involve heterogeneous biological reaction centers with immobilized cofactors and strong local spin interactions and are therefore not directly comparable to homogeneous liquid-state donor–acceptor systems. Consequently, systematic wavelength-resolved photo-CIDNP measurements in liquids remain relatively scarce. Existing studies typically compare only a limited number of discrete excitation wavelengths instead of continuously mapping the spectral dependence for a single molecular system. In addition, quantitative comparison across a broad spectral range requires normalization of the CIDNP response to the photon flux, which is not always implemented. As a result, the relationship between excitation wavelength, optical absorption, excited-state relaxation, and CIDNP efficiency in solution is not adequately understood.

In this work, we investigate excitation wavelength as an independent experimental parameter governing liquid-state photo-CIDNP. Using the tetraphenylporphyrin–benzoquinone (TPP–BQ) donor–acceptor system as a model system, we perform wavelength-resolved photo-CIDNP measurements across the spectral range of 350--800\,nm, and compare the resulting CIDNP action spectrum with the corresponding optical absorption profile. To ensure the robustness of the observed trends, CIDNP measurements were carried out using two complementary excitation approaches: tunable monochromatic laser excitation and spectrally selected broadband xenon-lamp illumination—and the CIDNP response was normalized to the photon flux at each wavelength. This approach enables a systematic assessment of how excitation wavelength influences photo-CIDNP efficiency and establishes excitation wavelength as a parameter that should be considered alongside other experimental variables in liquid-state photo-CIDNP studies.

\section{Framework for Interpreting Wavelength-Dependent Photo-CIDNP}

Interpretation of wavelength-resolved photo-CIDNP requires linking the observed nuclear spin polarization to the sequence of photophysical and spin-dependent processes that follow optical excitation in radical-pair systems \cite{SteinerUlrich1989_MagneticFieldEffects,ClossForbesNorris1987_SpinPolarizedEPR}. In liquid-state donor–acceptor systems, the CIDNP amplitude reflects not only the efficiency of photon absorption but also the probability that the absorbed excitation energy ultimately leads to the formation of spin-correlated radical pairs and observable nuclear spin polarization. In a simplified form, the CIDNP signal may be expressed as

\begin{equation}
CIDNP(\lambda) \propto N_{abs}(\lambda)\times \Phi_{RP}(\lambda)\times S(\lambda)\,,
\end{equation}

where $N_{abs}(\lambda)$ is the total number of photons absorbed by the sample during the irradiation period,

$\Phi_{RP}(\lambda)$ is the wavelength-dependent radical-pair yield, and 
$S(\lambda)$ represents the combined effects of radical-pair spin evolution, spin-selective reaction pathways, and nuclear-spin detection.

This relationship illustrates an important point: CIDNP is not expected to scale solely with optical absorption. Even when light at two wavelengths is absorbed with similar efficiency, it may populate distinct excited states or access alternative relaxation pathways, ultimately affecting radical-pair yields and the resulting CIDNP amplitudes.

Such considerations are particularly relevant for porphyrin-based donor–acceptor systems. Porphyrins exhibit structured absorption spectra consisting of several electronic bands, most prominently the intense Soret band in the near-UV/blue spectral region and the weaker Q-bands at longer wavelengths \cite{Gouterman1961_SpectraPorphyrins,Gust1991_ProtonAssistedPET}. Excitation within these bands may initially populate different excited-state manifolds, which subsequently relax through competing pathways including internal conversion, fluorescence, intersystem crossing, and electron transfer. The relative contributions of these processes can influence the efficiency of spin-correlated radical-pair formation and thus affect the resulting CIDNP signal. The interplay between these relaxation pathways may depend on the excitation wavelength and the CIDNP action spectrum is not necessarily expected to follow the optical absorption profile. Instead, wavelength-dependent changes in CIDNP can reflect variations in the balance between competing photophysical processes that precede radical-pair formation.

In the following, wavelength-resolved photo-CIDNP measurements are used to examine how the nuclear spin polarization of the TPP–BQ system varies across a broad spectral range and how this behavior compares with the corresponding optical absorption spectrum.

\section{Experimental Section}
\subsection{Chemical System and Sample Preparation}

The investigated system is based on the well-established photo-CIDNP reaction between tetraphenylporphyrin (TPP), acting as a photosensitizer and electron donor, and 1,4-benzoquinone (BQ), serving as an electron acceptor.
Upon photoexcitation, this donor-acceptor pair forms spin-correlated
radical pairs and has been widely employed as a model system for
investigating photo-CIDNP mechanisms in both high-field and low-field
NMR studies
\cite{Gust1991_ProtonAssistedPET,
sheberstov2021photochemically,
chuchkova2023magnetometer}.

Previous studies have shown that acidic conditions play an important
role in facilitating efficient CIDNP generation in porphyrin-quinone
systems
\cite{Gust1991_ProtonAssistedPET,
sheberstov2021photochemically,
chuchkova2023magnetometer}. Under these conditions, protonated forms of porphyrin and quinone species are expected to contribute significantly to the electron-transfer processes leading to the formation of spin-correlated radical pairs and nuclear spin polarization.

Stock solutions of TPP and BQ were prepared in chloroform. For the lamp-based experiments, the final concentrations were 0.5\,mM for TPP and 5\,mM for BQ in a solvent mixture of deuterated chloroform (CDCl$_3$) and acetic acid (CH$_3$COOH) (70:30, v/v). The solvent composition and concentration range were selected according to protocols established in our previous studies \cite{sheberstov2021photochemically,chuchkova2023magnetometer}. Both protonated and deuterated solvents were used depending on the experimental requirements.

For laser-based experiments and UV--Vis absorption measurements, the solvent composition and sample preparation procedure were kept unchanged, while lower concentrations of TPP and BQ were employed (0.03\,mM and 0.3\,mM, respectively). The lower concentrations were also used for the UV–Vis measurements to keep the absorbance within the reliable measurement range of the spectrophotometer and to avoid saturation of the intense Soret band. These concentrations were chosen to accommodate the different excitation geometry of the laser setup. In the lamp-based configuration, the sample was irradiated laterally through the wall of a standard 5\, mm NMR tube, while in the laser-based setup, the excitation light was delivered directly into the center of the sample volume via an optical fiber. The reduced concentrations therefore ensured suitable optical densities for efficient and homogeneous excitation under the corresponding experimental conditions.

Although the absolute CIDNP amplitudes depend on concentration and excitation geometry, the objective of the present work is not a direct comparison of signal amplitudes obtained in the two experimental configurations. Instead, the focus is placed on the wavelength dependence of the CIDNP response within each setup. As shown below, the resulting spectral trends are reproducible and remain consistent after normalization to the photon flux.

All samples were degassed by bubbling nitrogen gas for 1\,min prior to irradiation. The degassing procedure and duration were identical for all experiments.

\subsection{Optical Excitation and Wavelength-Scanning Setup}

A custom-built optical excitation system was developed to perform wavelength-resolved photo-CIDNP measurements over a broad spectral range. As the primary light source, a high-intensity broadband xenon lamp was employed. The lamp provides continuous optical emission from the near-ultraviolet to the near-infrared region (350--800\,nm), making it suitable for systematic excitation-wavelength studies.

\begin{figure}
\centering
\includegraphics[width=\linewidth]{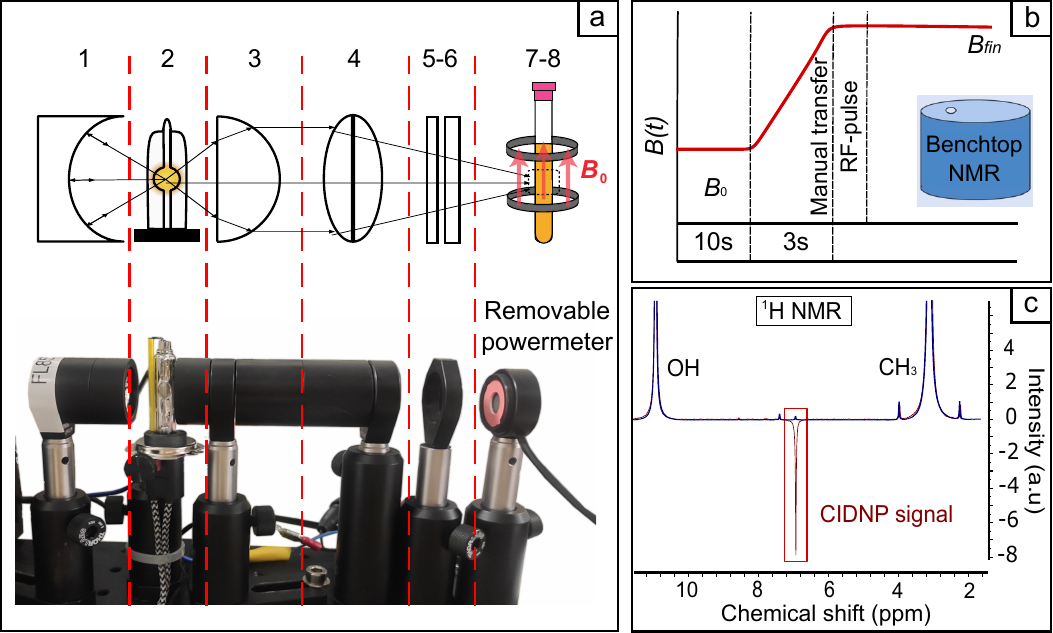}
\caption{Experimental concept and setup for wavelength-resolved photo-CIDNP measurements. (a) Schematic representation and photograph of the broadband lamp-based irradiation system. Optical elements are numbered as follows: (1) concave mirror, (2) broadband xenon lamp, (3) condenser lens, (4) focusing lens, (5--6) interchangeable optical filters for wavelength selection, (7--8) NMR tube and permanent magnet (50\,mT). The NMR tube containing the sample is positioned in a static magnetic field $B_0$ during optical irradiation. (b) Timing scheme of the photo-CIDNP experiment, illustrating the polarization period, manual transfer of the sample to the benchtop NMR spectrometer (3\,s), and subsequent radiofrequency excitation and signal acquisition. (c) Representative photo-CIDNP-enhanced $^1$H NMR signal obtained for the tetraphenylporphyrin/benzoquinone (TPP/BQ) system. The blue line shows the thermal-equilibrium spectrum and the red line the photo-polarized spectrum.}
\label{scemeLAMP}
\end{figure}

Excitation wavelength selection was achieved using a modular set of interchangeable 10-nm-FWHM band-pass interference filters (Figure\,\ref{scemeLAMP}a). This approach enables measurements across a broad spectral range without realignment of the optical system or repositioning of the sample. As a result, systematic variations during extended measurement series are minimized.

The filtered light was focused onto the NMR tube positioned within a low-field polarization magnet, ensuring reproducible irradiation conditions for all wavelengths. The broadband light source was a 30 W xenon lamp. Because the nominal lamp power refers to its total broadband output, it was not used for photon-flux normalization. Wavelength selection was performed using Thorlabs band-pass interference filters with a full width at half maximum of 10 nm. The optical power was measured separately for each filter directly at the sample position using a calibrated power meter. For these measurements, the NMR tube was removed and the power-meter sensor was placed at the same position normally occupied by the sample. The incident photon flux was calculated from the measured optical power and the central wavelength of the corresponding filter. Standard borosilicate-glass NMR tubes were used throughout the study. Borosilicate glass exhibits high optical transparency within the investigated spectral range and therefore does not introduce significant wavelength-dependent transmission losses under the present experimental conditions.

The overall experimental concept is summarized in Figure\,\ref{scemeLAMP}. Figure\,\ref{scemeLAMP}a shows the optical layout of the excitation system. Figure\,\ref{scemeLAMP}b illustrates the polarization and transfer sequence used in the photo-CIDNP experiments. Figure\,\ref{scemeLAMP}c presents a representative photo-CIDNP-enhanced $^1$H NMR spectrum of the TPP/BQ system.

For comparison, measurements were also performed using two laser sources.

One is a tunable optical parametric oscillator (OPO) laser system manufactured by HÜBNER Photonics. Measurements at 420 nm were performed using a TOPTICA DL pro diode laser. The specified laser linewidth for both systems was below 1 MHz. For the fiber-based configuration, the optical power was measured directly at the output of the optical fiber before its insertion into the sample.

\begin{figure}[!ph]
\centering
\includegraphics[width=0.85\linewidth]{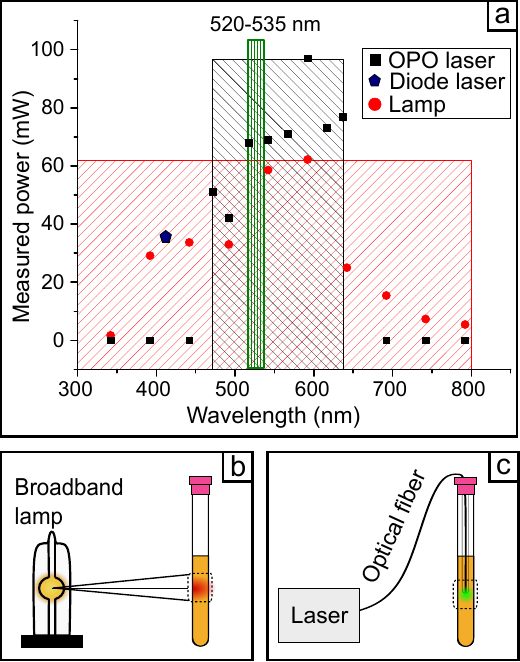}
\caption{(a) Experimentally measured optical power delivered to the sample
as a function of excitation wavelength for the xenon lamp (red circles),
OPO laser (black squares), and diode laser (blue pentagon). Shaded regions
indicate the wavelength ranges accessible under stable operating conditions.
The vertical green line marks the excitation wavelength region commonly
employed in conventional photo-CIDNP experiments (520--535 nm).
The overlap between lamp- and laser-based measurements enables direct
comparison of the two excitation approaches. (b,c) Schematic representations
of the lamp-based and laser-based polarization geometries, respectively.}
\label{powerLL}
\end{figure}

The experimentally accessible wavelength ranges and measured optical powers for both excitation schemes are summarized in Figure\,\ref{powerLL}a. The output power of the OPO varies substantially across the tuning range owing to wavelength-dependent phase-matching and resonator conditions. Consequently, the values shown in Figure\,\ref{powerLL}a correspond to the actual optical power delivered to the sample at each wavelength.

Measurements at 420\,nm were performed using a diode laser because this wavelength was not accessible under stable operating conditions of the OPO system. All remaining laser-based measurements were carried out using the same OPO configuration.

The two excitation schemes provide complementary capabilities. The laser source offers narrow-band excitation and high power density within a restricted wavelength range. The xenon-lamp setup enables measurements across a substantially broader spectral region. The overlap between the two systems allows direct comparison of wavelength-dependent CIDNP measurements obtained using different excitation approaches.

Schematic representations of both polarization configurations are shown in Figure\,\ref{powerLL}b,c.

\subsection{NMR Detection and Measurement Logic}

NMR detection was performed using a Magritek Spinsolve 60 benchtop spectrometer. The experimental protocol was designed to detect photo-CIDNP-generated nuclear spin polarization while minimizing polarization losses prior to signal acquisition. In all experiments, the sample was first irradiated for a defined polarization period and the resulting non-equilibrium nuclear magnetization was subsequently detected by NMR. The obtained spectra were compared with corresponding dark-reference measurements acquired under identical conditions without optical excitation.

For the lamp-based experiments, nuclear spin polarization was generated in a dedicated low-field permanent magnet operating at 50\,mT. The irradiation time was fixed at 10\,s. After the polarization period, the sample was manually transferred to the benchtop NMR spectrometer, resulting in a transfer delay of approximately 3\,s (Figure\,\ref{scemeLAMP}b). This configuration was dictated by the optical geometry of the broadband excitation setup and was used consistently throughout all wavelength-dependent lamp measurements.

For the laser-based experiments, the sample remained inside the NMR detection magnet during optical irradiation. Excitation light was delivered directly into the sample volume through an optical fiber (Figure\,\ref{powerLL}c), allowing irradiation and detection to be performed without sample transfer. Under these conditions, a polarization time of 5\,s was sufficient to obtain reproducible photo-CIDNP signals.

The irradiation times employed in the two experimental configurations were optimized independently according to the corresponding excitation geometry and available optical power. Since the present work focuses on the wavelength dependence of the photo-CIDNP response within each excitation scheme, direct comparison of absolute signal amplitudes between lamp- and laser-based measurements is not intended.

The observed photo-CIDNP signals originate predominantly from benzoquinone protons. Comparison of the polarized benzoquinone signals with the thermally polarized solvent signals allows estimation of the signal enhancement factor and the corresponding nuclear spin polarization. Under optimized experimental conditions, photo-CIDNP can generate nuclear spin polarization several orders of magnitude greater than thermal equilibrium, leading to a substantial enhancement of the NMR signal \cite{MorozovaIvanov2019_TimeResolvedCIDNPBiomolecules}.

\subsection{Wavelength Scanning Protocol}

For each excitation wavelength, the sample was irradiated for a fixed polarization period determined by the corresponding experimental configuration (5\,s for laser-based measurements and 10\,s for lamp-based measurements). These irradiation times were kept constant throughout each measurement series to ensure consistent experimental conditions.

To eliminate possible cumulative photochemical effects and ensure identical initial conditions, a freshly prepared sample was used for each excitation wavelength. All samples were prepared according to the protocol described above. For each wavelength point, five consecutive photo-CIDNP measurements were performed on the same sample. Between successive measurements, a delay of approximately 3\,min was introduced to allow complete relaxation of the nuclear magnetization back to thermal equilibrium before the next irradiation cycle. This procedure ensured that each measurement started from identical initial spin conditions while minimizing the influence of residual polarization from the preceding experiment.

The reported CIDNP amplitudes correspond to mean values obtained from the five repeated measurements, while the error bars represent one standard error. All NMR spectra were recorded using identical acquisition parameters within a given experimental configuration. Dark-reference measurements were performed periodically to verify sample stability in the absence of illumination.

\section{Results}

\subsection{Wavelength dependence of the photo-CIDNP signal}

\begin{figure*}
    \centering
    \includegraphics[width=.7\linewidth]{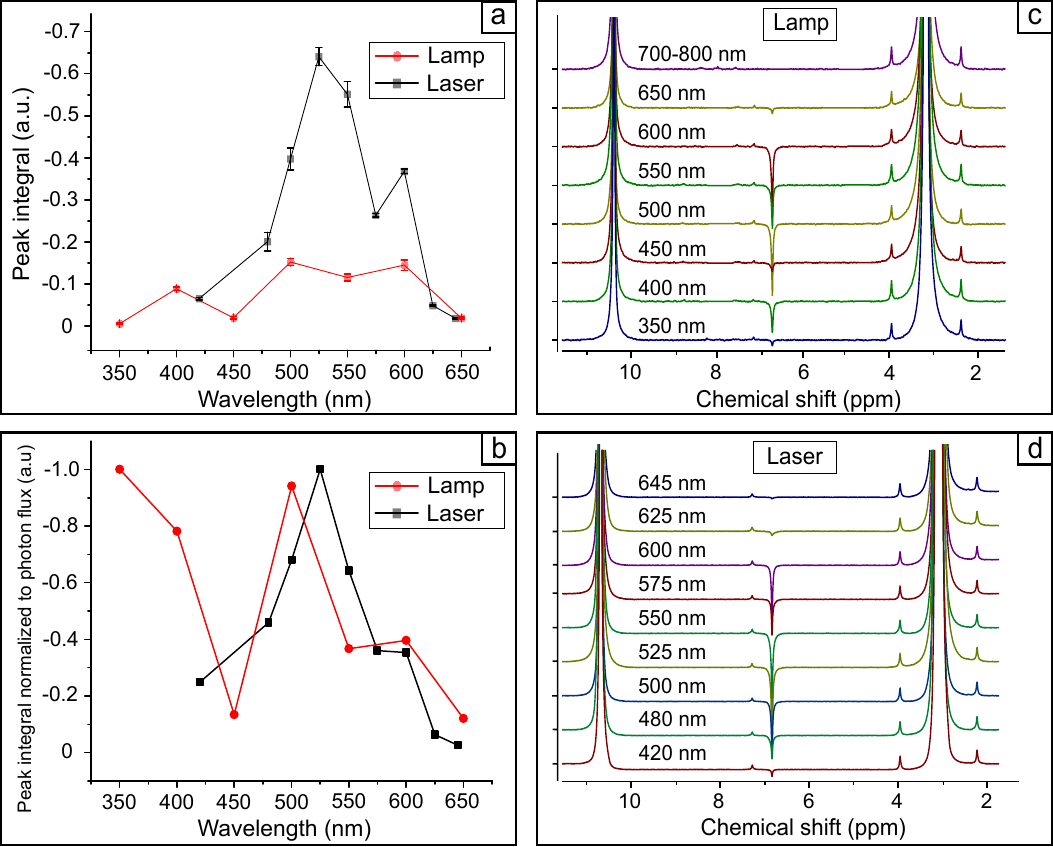}
    \caption{Comparison of wavelength-dependent photo-CIDNP signals obtained using broadband lamp and laser excitation. (a) Integrated photo-CIDNP signal intensity as a function of excitation wavelength. Red circles correspond to measurements performed using a broadband lamp with interference filters, whereas black squares represent measurements obtained using an OPO/diode laser system. (b) The same data normalized to the incident photon flux and irradiation time, with each dataset subsequently normalized to its maximum value. (c,d) Representative $^1$H NMR photo-CIDNP spectra recorded at selected excitation wavelengths using the lamp-based (10 s irradiation time) and laser-based (5 s irradiation time) setups.
}
    \label{curvesINTvsWL.pdf}
\end{figure*}

The wavelength dependence of the photo-CIDNP response obtained using broadband lamp and laser excitation is summarized in Figure\,\ref{curvesINTvsWL.pdf}. Figure\,\ref{curvesINTvsWL.pdf}a presents the integrated CIDNP signal intensity as a function of excitation wavelength, where red circles correspond to broadband lamp excitation and black squares to laser excitation. Figure\,\ref{curvesINTvsWL.pdf}b shows the same data after normalization to the incident photon flux. Representative $^1$H photo-CIDNP spectra recorded at selected wavelengths are shown in Figures\,\ref{curvesINTvsWL.pdf}c and d.

The raw CIDNP amplitudes shown in Figure\,\ref{curvesINTvsWL.pdf}a differ substantially between the lamp- and laser-based experiments. Such differences are expected because the measurements were performed under different experimental conditions, including excitation power, sample concentration, polarization magnetic field, excitation geometry, and illuminated sample volume. As a result, direct comparison of the absolute signal amplitudes between the two excitation schemes is not particularly meaningful. Both datasets exhibit similar wavelength-dependent behavior. In particular, a pronounced CIDNP response is observed in the near-ultraviolet region around 350\,nm, followed by a substantial decrease in the 400--450\,nm range and a subsequent increase toward a broad maximum in the absolute CIDNP amplitude centered near 500\,nm. At longer wavelengths, the CIDNP signal gradually decreases and becomes negligible above approximately 650--700\,nm. These trends are directly reflected in the representative spectra shown in Figures\,\ref{curvesINTvsWL.pdf}c and d. Although the absolute signal amplitudes vary significantly with excitation wavelength, the characteristic changes observed in the integrated CIDNP intensities are correlated in the corresponding NMR spectrum.

The experimental uncertainties shown in Figure\,\ref{curvesINTvsWL.pdf} are generally small compared to the overall wavelength-dependent variations. A somewhat larger scatter is observed in the 480--550\,nm region, where the CIDNP response reaches its maximum values. Although the present measurements do not allow quantitative conclusions regarding polarization build-up kinetics, this observation may indicate an increased sensitivity of the CIDNP amplitude to small variations in irradiation conditions within this spectral range.

An additional experimental observation is the appearance of visible luminescence during irradiation in the 400--450\,nm spectral region. For the lamp-based measurements, luminescence was clearly observed at 400 and 450\,nm, while for the laser-based experiments it was observed at 420\,nm. Interestingly, the CIDNP signal remains relatively pronounced at 400\,nm but decreases substantially at longer wavelengths within this spectral region, c.f. detailed discussion below.

\subsection{Cross-validation of the wavelength dependence using two excitation schemes}

A more meaningful comparison between the two excitation schemes is obtained after normalization of the CIDNP amplitudes to the incident photon flux and irradiation time. The normalized wavelength dependence is shown in Figure\,\ref{curvesINTvsWL.pdf}b, where red circles correspond to the broadband lamp measurements and black squares to the laser-based measurements.

Despite substantial differences in experimental conditions, including excitation geometry, optical power density, sample concentration, polarization magnetic field, and illuminated sample volume, both datasets exhibit remarkably similar wavelength-dependent behavior. In particular, both excitation schemes reproduce the same characteristic features: enhanced CIDNP activity in the near-ultraviolet region, a pronounced reduction of the signal in the 400--450\,nm  range, and a broad region of increased CIDNP efficiency centered near 500\,nm . 
The qualitative agreement between the two datasets suggests that the main spectral features are robust with respect to the excitation source and experimental configuration.
Instead, the data suggest that the wavelength dependence reflects intrinsic properties of the photo-CIDNP process in the TPP/BQ system.

However, minor deviations between the lamp- and laser-based measurements remain visible, especially in the near-ultraviolet region and near the signal minimum. Such differences arise from the finite bandwidth of the optical filters used in the lamp setup, differences in spatial illumination profiles, and the laser measurements sample only discrete wavelengths. In addition, the 420\,nm  data point was obtained using a separate diode laser source because this wavelength was not accessible with sufficient output power from the OPO system.

Overall, the close correspondence between the normalized datasets provides an important internal consistency check of the experimental approach. The agreement between two independent excitation schemes supports the conclusion that excitation wavelength acts as a genuine experimental parameter influencing photo-CIDNP efficiency and motivates a more detailed comparison with the optical absorption properties of the system.

\subsection{Correlation between optical absorption and CIDNP efficiency}

\begin{figure}
    \centering
    \includegraphics[width=0.8\linewidth]{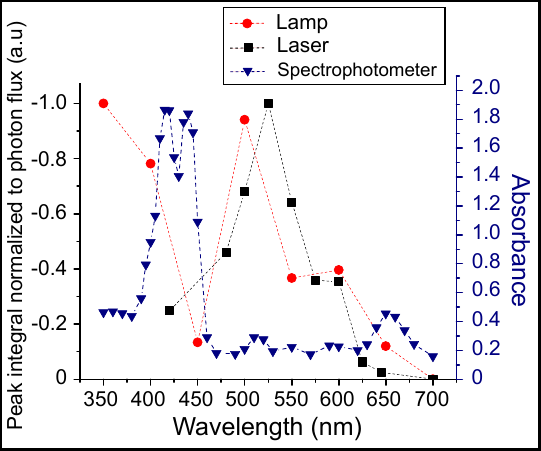}

    \caption{Comparison of the wavelength dependence of the photon-flux-normalized
photo-CIDNP signal intensity measured using broadband lamp excitation
(red circles) and diode/OPO laser excitation (black squares) with the optical
absorption spectrum of the sample obtained by UV--Vis spectroscopy
(blue inverted triangles, right axis).}
    \label{SpectrophotomVSnorm}
\end{figure}

To examine the relationship between optical absorption and CIDNP efficiency, the photon-flux-normalized CIDNP signal was compared with the UV--Vis absorption spectrum of the sample (Figure\,\ref{SpectrophotomVSnorm}). In this representation, the normalized CIDNP amplitudes reflect the efficiency of CIDNP generation per incident photon flux, whereas the absorption spectrum characterizes the wavelength dependence of light absorption by the sample.

Figure\,\ref{SpectrophotomVSnorm} shows the discrepancy between the absorption profile and the CIDNP response, i.e. the absorption spectrum exhibits its strongest features in the 400--440\,nm region, while the absolute magnitude of the normalized CIDNP signal reaches a pronounced minimum in the same spectral range. 
Conversely, enhanced CIDNP activity is observed near 350\,nm  and around 500\,nm, where the absorption spectrum does not display correspondingly dominant maxima. An additional indication that CIDNP efficiency is not governed solely by optical absorption is provided by the behavior within the 400--450\,nm  region itself. The UV--Vis spectrum shows nearly identical absorbance values at 400 and 450\,nm

(approximately 1.1\,a.u.), whereas the corresponding normalized CIDNP amplitudes differ substantially. Thus, even within a spectral range where the probability of photon absorption remains similar, the efficiency of CIDNP generation varies markedly.

A further experimental observation is the appearance of strong visible luminescence during excitation in the 400--450\,nm  region. Luminescence was observed for lamp-based excitation at 400 and 450\,nm  and for laser excitation at 420\,nm, coinciding with the spectral region where the CIDNP response is strongly reduced. Interestingly, the CIDNP signal remains relatively pronounced at 400\,nm  but decreases substantially at longer wavelengths within this region. While the present data do not allow a direct determination of the underlying relaxation mechanisms, the correlation between enhanced luminescence and reduced CIDNP efficiency suggests that wavelength-dependent excited-state processes may contribute significantly to the observed CIDNP response.

Taken together, these observations demonstrate that the CIDNP action spectrum contains information beyond the optical absorption profile alone. The pronounced variations observed within the 350--500\,nm  region indicate that processes occurring after photon absorption play an important role in determining the final CIDNP efficiency. Possible interpretations of these observations are discussed in the following section.

\section{Discussion}

\subsection{Possible origins of the wavelength dependence}

\begin{figure}
    \centering
    \includegraphics[width=0.75\linewidth]{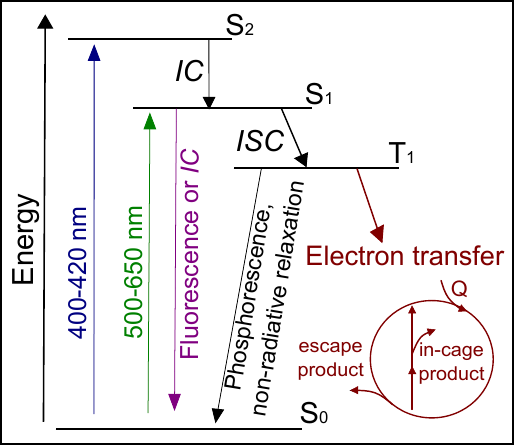}
    \caption{Schematic Jablonski diagram illustrating the photophysical and photochemical pathways relevant to photo-CIDNP in a system containing tetraphenylporphyrin (TPP) and benzoquinone (BQ). Optical excitation in the 400--420~nm range populates higher singlet states ($\mathrm{S}_2$), followed by rapid internal conversion ($IC$) to $\mathrm{S}_1$, while excitation in the 500--650~nm region directly addresses $\mathrm{S}_1$. From $\mathrm{S}_1$, the system can undergo fluorescence, internal conversion to $\mathrm{S}_0$, or intersystem crossing ($ISC$) to the triplet state $\mathrm{T}_1$. The triplet state relaxes via phosphorescence or non-radiative decay. Electron transfer to the quencher (Q) forms a radical pair, resulting in CIDNP through recombination or escape pathways.}
    \label{energyLV}
\end{figure}

A schematic energy-level diagram summarizing the proposed interpretation of the wavelength-dependent CIDNP response is shown in Figure\,\ref{energyLV}. The comparison between the CIDNP action spectrum and the UV--Vis absorption spectrum demonstrates that optical absorption alone cannot account for the observed wavelength dependence of the photo-CIDNP response. In particular, excitation wavelengths associated with similar absorption coefficients produce substantially different CIDNP amplitudes, indicating that processes occurring after photoexcitation play a decisive role in determining the final nuclear spin polarization.

Within the radical-pair framework, CIDNP arises from spin-selective reactions of radical pairs formed following photoinduced electron transfer \cite{KapteinOosterhoff1969_CIDNP_II,SteinerUlrich1989_MagneticFieldEffects,HoreBroadhurst1993_PhotoCIDNP}. Consequently, the observed wavelength dependence reflects not only the probability of photon absorption but also the competition between excited-state relaxation pathways that either promote or suppress radical-pair formation.

Excitation in the Q-band region around 500--600\,nm directly populates the lowest excited singlet state and is associated with efficient CIDNP generation. In contrast, excitation in the Soret region initially populates higher-lying singlet states, which subsequently relax through a combination of internal conversion, fluorescence, intersystem crossing, and electron-transfer processes. Variations in the relative contributions of these competing pathways can substantially alter the efficiency of radical-pair formation and therefore the resulting CIDNP response.

The experimentally observed luminescence in the 400--450\,nm region is consistent with this interpretation. Strong visible luminescence was observed for lamp-based excitation at 400 and 450\,nm and for laser excitation at 420\,nm, coinciding with the spectral region where the normalized CIDNP signal reaches a pronounced minimum. Although the present data do not allow direct identification of the underlying relaxation mechanism, the correlation between enhanced luminescence and reduced CIDNP efficiency suggests that a larger fraction of the excited-state population relaxes through pathways that do not contribute effectively to radical-pair formation.

Particularly intriguing is the enhanced CIDNP response observed at the shortest investigated wavelengths. The strong polarization detected near 350\,nm cannot be readily explained by the absorption spectrum alone and suggests that additional excited-state processes become important in the near-UV region. Recent theoretical \cite{aydin2016geometric} study of tetraphenylporphyrin and related porphyrin derivatives predict the presence of energetically close singlet and triplet excited states, including several singlet--triplet crossings near 393\,nm and at higher excitation energies. Such crossings may provide additional relaxation pathways from higher excited singlet states into triplet manifolds prior to complete internal conversion. If present, these pathways could modify the balance between fluorescence, intersystem crossing, and electron transfer, thereby influencing the yield of spin-correlated radical pairs and ultimately the CIDNP efficiency.

The precise energetic positions of these states are expected to depend sensitively on the molecular environment. Solvent composition, protonation equilibria, and possible coexistence of different porphyrin species may alter the relative energies of singlet, triplet, and charge-transfer states, thereby modifying the efficiency of competing relaxation pathways\cite{aydin2016geometric,aydin2014comparative,lan2007absorption,SteinerUlrich1989_MagneticFieldEffects}. In this regard, it is noteworthy that the UV--Vis spectrum exhibits both a broadened Soret region and enhanced long-wavelength absorption extending toward approximately 670\,nm\cite{aydin2016geometric}. Such spectral features indicate the coexistence of more than one porphyrin species in solution and are qualitatively consistent with partial protonation of tetraphenylporphyrin under acidic conditions. Although the available data do not permit a definitive assignment, even a minor population of protonated species could influence excited-state relaxation dynamics and contribute to the observed wavelength dependence of CIDNP generation. At the same time, the present results do not allow a unique mechanistic interpretation of the enhanced near-UV response. Several factors—including higher excited states, singlet--triplet crossings, solvent-induced energetic shifts, protonation effects, and charge-transfer-state energetics—may contribute simultaneously. Furthermore, access to the shortest wavelengths was experimentally limited by the available excitation sources and the relatively low optical power in the near-UV region, preventing denser spectral sampling around the observed enhancement. Consequently, the current data should be viewed primarily as evidence that the 350--500\,nm region exhibits particularly rich wavelength-dependent behavior rather than as proof of a specific microscopic mechanism.

Taken together, the results indicate that excitation wavelength acts as a genuine control parameter of the photo-CIDNP process by influencing the balance between competing excited-state relaxation pathways. At the same time, the observations identify the near-UV and blue spectral regions as particularly promising targets for future experimental and theoretical investigations aimed at understanding how excited-state dynamics influence radical-pair generation and CIDNP efficiency.

\subsection{Advantages and limitations of lamp-based and laser-based excitation}

The lamp-based and laser-based excitation schemes employed in this work provide complementary approaches for wavelength-resolved photo-CIDNP studies. Laser excitation offers high optical power density, narrow spectral bandwidth, and direct irradiation of the sample inside the NMR spectrometer, enabling efficient polarization and immediate signal detection without sample transfer. However, laser systems are often limited by incomplete wavelength coverage, wavelength-dependent power variations, and higher experimental complexity.

In contrast, the lamp-based setup provides broad spectral accessibility using a single excitation source and allows rapid wavelength selection through interchangeable optical filters. The system is compact, portable, readily adaptable to different molecular systems, and can be implemented using relatively inexpensive and widely available optical components. Its main limitations are the lower optical power density and reduced spectral selectivity compared to laser excitation. Despite substantial differences in excitation geometry, magnetic field, sample concentration, and optical power, both approaches reproduce the same overall wavelength dependence of the CIDNP response after photon-flux normalization. This qualitative agreement confirms that the observed spectral trends are intrinsic to the photo-CIDNP process rather than specific to a particular experimental configuration.

Taken together, these results highlight the utility of spectrally
filtered broadband non-laser sources as cost-effective and versatile
platforms for liquid-state photo-CIDNP, complementing established
light-coupled NMR implementations based on LEDs and continuous
illumination
\cite{Feldmeier2013_LEDIllumination,
sheberstov2021photochemically,
chuchkova2023magnetometer}.

The two excitation schemes should therefore be viewed as
complementary rather than competing approaches. Their combined use
provides an independent validation of the wavelength dependence of
photo-CIDNP and demonstrates that systematic wavelength-resolved
measurements can be implemented using either platform, depending on
the required spectral coverage, optical power density, and
experimental geometry.

\section{Conclusions}

In this work, we systematically investigated the excitation-wavelength dependence of liquid-state photo-CIDNP in the tetraphenylporphyrin–benzoquinone donor–acceptor system across a broad spectral range extending from the near-ultraviolet to the near-infrared. By combining broadband lamp-based excitation with tunable laser measurements and applying photon-flux normalization, we obtained a wavelength-resolved CIDNP action spectrum under a wide range of excitation conditions.

The results demonstrate that the efficiency of photo-CIDNP cannot be predicted solely from the optical absorption spectrum of the sample. Significant differences in CIDNP efficiency were observed across the investigated spectral range, including a pronounced suppression of the polarization response in the 400--450\,nm region and enhanced CIDNP activity near 350\,nm and around 500\,nm. These observations indicate that excitation wavelength influences not only the probability of photon absorption but also the subsequent excited-state processes that govern radical-pair formation and nuclear spin polarization.

Comparison of the lamp-based and laser-based measurements revealed a high degree of agreement after normalization to the photon flux, despite substantial differences in excitation geometry, optical power, magnetic field, and sample concentration. This agreement confirms that the observed wavelength dependence represents an intrinsic property of the photo-CIDNP process rather than an artifact of a particular experimental configuration. At the same time, the results demonstrate that both broadband and laser-based excitation schemes can be successfully employed for systematic wavelength-resolved photo-CIDNP studies.

The strongest deviations from a simple absorption-driven behavior were observed in the 350--500\,nm spectral region, where relatively small changes in excitation wavelength produced substantial variations in CIDNP efficiency. Although the present data do not permit a definitive mechanistic interpretation, they suggest that excited-state relaxation pathways play a central role in determining the efficiency of CIDNP generation. This spectral region therefore represents a particularly promising target for future experimental and theoretical investigations.

Several important questions remain open. In particular, future studies should address the influence of excitation wavelength under varying solvent compositions, protonation conditions, and donor–acceptor concentrations in order to disentangle the respective contributions of molecular environment and excited-state dynamics. Higher-resolution measurements in the near-UV and blue spectral regions, combined with time-resolved spectroscopic approaches and theoretical modeling, will be necessary to clarify the origin of the observed wavelength dependence and the mechanisms responsible for the enhanced CIDNP response at short excitation wavelengths.

More broadly, the present work establishes excitation wavelength as an independent and experimentally accessible control parameter in liquid-state photo-CIDNP. A deeper understanding of wavelength-dependent spin polarization may contribute not only to the elucidation of fundamental spin-chemical processes but also to the rational optimization of photo-CIDNP as a hyperpolarization technique\cite{eills2023spin}. In the long term, such knowledge may facilitate the development of more efficient photo-CIDNP methodologies for chemically and biologically relevant systems, including studies of dilute molecular targets where sensitivity enhancement remains a critical challenge.

\balance

\bibliography{CIDNP_references.bib} 
\bibliographystyle{rsc} 
\end{document}